\documentclass[aps,prc,floatfix]{revtex4}
\usepackage{amsmath}
\usepackage{bm}
\usepackage{psfig}
\usepackage{graphicx}
\usepackage{dcolumn}

\begin{document}

\newcommand{\etal}{{\it et al.\/ }}
\newcommand{\GeVsq}{(GeV/$c$)$^2$}

\title{RDWIA analysis of $^{12}$C$(e,e^\prime p)$ for 
$Q^2 < 2$ (GeV/$c$)$^2$}
\author{James J. Kelly}
\affiliation{Department of Physics, University of Maryland, 
College Park, MD 20742}
\date{January 28, 2005}

\begin{abstract}
We analyze data for $^{12}$C$(e,e^\prime p)$ with $Q^2 < 2$ (GeV/$c$)$^2$
using the relativistic distorted-wave impulse approximation (RDWIA) based
upon Dirac-Hartree wave functions.
The $1p$ normalization extracted from data for $Q^2 > 0.6$ (GeV/$c$)$^2$ is
approximately 0.87, independent of $Q^2$, which is consistent with the
predicted depletion of the $1p_{3/2}$ orbital by short-range correlations.
The total $1p$ and $1s$ strength for $E_m < 80$ MeV approaches
100\% of IPSM, consistent with a continuum contribution for 
$30< E_m < 80$ MeV of about 12\% of IPSM.
Similarly, a scale factor of 1.12 brings RDWIA calculations into good
agreement with $^{12}$C$(e,e^\prime p)$ data for transparency.
We also analyzed low $Q^2$ data from which a recent NRDWIA analysis
suggested that spectroscopic factors might depend strongly upon the
resolution of the probe.
We find that the momentum distributions for their empirical Woods-Saxon 
wave functions fit to low $Q^2$ data for parallel kinematics are too
narrow to reproduce data for quasiperpendicular kinematics, especially
for larger $Q^2$, and are partly responsible for reducing fitted
normalization factors.
Although the RDWIA normalization factors for $Q^2 < 0.2$ (GeV/$c$)$^2$ 
are also smaller than obtained for $Q^2 > 0.6$ (GeV/$c$)$^2$, the effect 
is smaller and we argue that it should be attributed to the effective
single-nucleon current operator instead of to spectroscopic factors 
which are probe-independent properties of nuclear structure.
However, remediation of the failure of RDWIA calculations to reproduce 
low $Q^2$ data for parallel kinematics will require a more sophisticated 
modification of the current than a simple multiplicative factor.
\end{abstract}

\maketitle

\section{Introduction}

It is generally believed that single-nucleon electromagnetic knockout 
provides unambiguous measurements of the momentum distributions and 
spectroscopic factors for single-hole states near the Fermi surface.
Reviews of analyses based upon the nonrelativistic distorted-wave
impulse approximation (NRDWIA) can be found in Refs. 
\cite{Lapikas93,Kelly96,Boffi96}, which show that the momentum 
distributions are described well by mean-field calculations, such as
Skyrme-Hartree-Fock, while spectroscopic factors for low-lying state
are reduced relative to the independent particle shell model (IPSM) 
by an average factor of about $65\%$ over a broad range of $A$.
Part of the depletion of IPSM strength can be attributed to 
short-range correlations that shift approximately $15\%$ of the
hole strength to very large missing energies, beyond 100 MeV.
Recently, a direct measurement of the spectral function using
$^{12}$C$(e,e^\prime p)$ in parallel kinematics \cite{Rohe04}
observed approximately 0.6 protons in a region with 
$p_m \gtrsim 240$ MeV/$c$ and $E_m \gtrsim 50$ MeV 
attributable to single-nucleon knockout from correlated clusters.
This finding is consistent with the $16\%$ IPSM predicted by 
Frick \etal \cite{Frick04} 
using the self-consistent Green's function (SCGF) method \cite{Dickhoff04}.
Similar estimates of the depletion of hole states are available
from the correlated basis function (CBF) theory \cite{Benhar90}
and other methods also.
Furthermore, SCGF calculations \cite{Muther94,Muther95a} show that the 
momentum distribution for low-lying quasihole states is hardly affected 
by short-range correlations, remaining very similar to mean-field 
calculations, while the momentum distributions for the large-$E_m$ 
continuum are substantially broadened.

If the occupancy of IPSM orbitals is predicted to be approximately $85\%$
by theories that reproduce the correlated strength at large $(E_m,p_m)$,
why is only $65\%$ observed by $(e,e^\prime p)$ experiments?
Coupling to collective modes is expected to produce significant fragmentation 
of the valence quasi-hole strength spread over a range of perhaps 10 MeV, 
but the extended RPA calculations  presently available do not yet reproduce 
those fragmentation patterns well \cite{Dickhoff04}.
Thus, it is possible that many small fragments are missed experimentally
even if they lie within the experimental acceptance.
Alternatively, problems in the reaction model may lead to systematic
underestimation of spectroscopic factors.
Most of the data contributing to the aforementioned estimate of
$65\%$ IPSM were limited to $T_p < 100$ MeV and were analyzed using
NRDWIA calculations based upon empirical Woods-Saxon wave functions.
More recent relativistic analyses (RDWIA) typically produce larger
spectroscopic factors \cite{Udias95} and describe interference response 
functions more accurately \cite{Kelly99b,Udias99,Udias00}.
The effects due to dynamical enhancement of Dirac spinors by the 
nuclear mean field are actually stronger at low $Q^2$ than at higher
energies \cite{Kelly05a}. 

While spectroscopic factors are of obvious importance to theories of nuclear
structure, they also affect analyses of nuclear transparency that look
for the onset of color transparency \cite{Dutta03,Garrow02}.
Those experiments typically compare the $(e,e^\prime p)$ yield
integrated over wide but finite ranges of missing momentum, $p_m$, and 
missing energy, $E_m$, with calculations based upon a model spectral function.
This spectral function is usually based upon the IPSM with a correction 
factor for the depletion by correlations of the single-hole strength within 
the experimental acceptance.
Therefore, the accuracy of the depletion factor affects the accuracy
of transparency measurements.

Recently Lapik\'as \etal \cite{Lapikas00} analyzed the data for
$^{12}$C$(e,e^\prime p)$ with $Q^2 < 0.4$ (GeV/$c$)$^2$ using NRDWIA 
calculations with Woods-Saxon bound-state wave functions fit to the NIKHEF 
data for $T_p = 70$ MeV in parallel kinematics \cite{vdSteenhoven88a} and
concluded that the normalization factors relative to IPSM were
$N_{1p} = 0.56 \pm 0.02$ and $N_{1s} = 0.59 \pm 0.04$ for
$E_m < 80$ MeV.
They then argued that the transparency results based upon an estimated 
depletion factor of $f = 0.9$ for the single-hole spectral function with 
$E_m < 80$ MeV should be increased by a factor of approximately 1.25 using
appropriately weighted attenuation factors for the $1p$ and $1s$ shells.
Finally, they argued that the normalization factors for $^{12}$C vary
strongly with $Q^2$ from the low values fitted to the NIKHEF data to
values approaching the IPSM limit for $Q^2 > 2$ (GeV/$c$)$^2$.
Frankfurt \etal \cite{Frankfurt01} speculated that quenching of
spectroscopic factors for low $Q^2$ might be caused by probing a 
dressed quasiparticle using long wave lengths and that at higher 
resolution the effective current operator for the $(e,e^\prime p)$ 
reaction approaches more closely that for a free nucleon. 
However, although the $Q^2$ range was smaller, our recent relativistic 
analysis showed no significant slope in the normalization factors fit to 
$^{16}$O$(e,e^\prime p)$ data for $Q^2 < 0.8$ (GeV/$c$)$^2$ \cite{Fissum04}.
Furthermore, we find that RDWIA calculations based upon the ordinary 
single-nucleon current with dynamical enhancement of lower components 
of Dirac spinors describe the left-right asymmetry for quasiperpendicular 
kinematics very well \cite{Kelly05a};
it is not obvious that one should describe quasiparticles for low $Q^2$
using Dirac spinors in a similar mean field.

In this paper we use relativistic distorted wave (RDWIA) calculations to
analyze the data for $^{12}$C$(e,e^\prime p)$ with 
$Q^2 \lesssim 2$ (GeV/$c$)$^2$.
In addition to the data considered by  Lapik\'as \etal \cite{Lapikas00},
we include recent data from Dutta \etal \cite{Dutta03} for $Q^2 = 0.6$,
1.2, and 1.8 (GeV/$c$)$^2$.
Our RDWIA fits produce systematically larger normalization factors 
using Dirac-Hartree wave functions than using the Woods-Saxon wave 
functions that were fit to the NIKHEF data.
The latter do not describe the momentum distributions for larger $Q^2$
well; nor do they describe the data for low $Q^2$ with quasiperpendicular
kinematics as well as do Dirac-Hartree wave functions.
Therefore, the fits to data for parallel kinematics with small ejectile
energies appear to produce artifically narrow momentum distributions 
that result in anomalously low spectroscopic factors.

The model and fitting procedures are described in Sec. \ref{sec:model}.
Sec. \ref{sec:EMA} presents an analysis based upon the effective 
momentum approximation (EMA) that permits direct comparison between 
nonrelativistic and relativistic bound-state wave functions.
Sec. \ref{sec:RDWIA} presents a more rigorous analysis, without
using the EMA, of both spectroscopic factors and transparency.
Our conclusions are summarized in Sec. \ref{sec:conclusions}.

\section{Model and Data}
\label{sec:model}

Detailed descriptions of RDWIA for $(e,e^\prime p)$ reactions are been
given in many recent papers.
We will use the methods and terminology of Refs. \cite{Fissum04,Kelly05a}.
All calculations treat electron distortion in the $q_\text{eff}$ 
approximation and use the CC2 current operator with MMD form factors 
\cite{MMD} in Coulomb gauge.
Optical potentials for Dirac phenomenology were taken from the 
global analysis by Cooper \etal \cite{Cooper93}.
We consider Dirac-Hartree wave functions from the original Horowitz and
Serot (HS) analysis \cite{Horowitz86} and NLSH wave functions from
Sharma \etal \cite{Sharma93}.
In addition we consider Woods-Saxon wave functions fit to 
$^{12}$C$(e,e^\prime p)$ data.

The data we consider are summarized in Table \ref{table:data}.
Reduced cross sections for the Tokyo, Saclay, and SLAC data were 
provided by Lapik\'as \cite{Lapikas00,Lapikas04} and include small 
adjustments to a common convention for reduced cross section.
In addition, radiative corrections were applied to the data from 
Ref. \cite{Bernheim82}.
Only for the NIKHEF experiment \cite{vdSteenhoven88a} was it possible to 
resolve the lowest three $1p$ fragments.
Although there are small differences between the missing momentum
distributions for these three fragments, the ground state represents
about $81\%$ and the next $1p_{3/2}$ fragment an additional $9\%$ of 
the total $1p$ strength for $E_m < 25$ MeV.
Therefore, we represent the $1p$ strength using either a single
Dirac-Hartree wave function for $1p_{3/2}$ or the NIKHEF fit to the
ground state.
That experiment was also able to resolve several weak positive-parity
states with $E_m < 30$ MeV, but their strength is small enough to
neglect in the present analysis \cite{vdSteenhoven88b}.
For most other experiments the lowest missing energy bin contains 
a small contribution from the low-energy tail of the $1s$ shell.
Furthermore, for all experiments the bin intended to emphasize the
$1s$ shell inevitably contains contributions from a broad continuum 
that may include additional $1p$ strength.
Therefore, each $E_m$ bin will be fit as an incoherent mixture of
$1p$ and $1s$ contributions according to
\begin{equation}
\label{eq:mixture}
\sigma_\text{red} = 
N_{1p} \sigma_\text{red}(1p_{3/2}) +  N_{1s} \sigma_\text{red}(1s_{1/2})
\end{equation}
where the overlap function for each shell is assumed to be
independent of missing energy and is normalized to full occupancy.
In addition, for the JLab data sets we fit a normalization factor 
for the independent particle shell model (IPSM) to the inclusive 
data in the range $10 \leq E_m \leq 80$ MeV according to
\begin{equation}
\label{eq:NIPSM}
\sigma_\text{red} = N_\text{IPSM} 
\left( \sigma_\text{red}(1p) +  \sigma_\text{red}(1s) \right)
\end{equation}
Calculations will be shown for the entire $p_m$ range measured for
each data set, but only data in the range $|p_m| \leq 200$ MeV$c$ will 
used to fit the normalization factors because larger $p_m$ may
be susceptible to increasing corrections for effects not contained
in the RDWIA model based upon single-nucleon current operators.

We also require a model of the $1s$ energy distribution to obtain the 
spectroscopic factor from the normalization 
$N_{1s}(E_\text{min},E_\text{max})$ fit to limited and variable intervals 
of $E_m$.
One often employs a Lorentzian distribution
\begin{equation}
\label{eq:L}
L_{1s}(E_m) = \frac{1}{2\pi} \frac{\Gamma(E_m)}
{(E_m-E_{1s})^2 + (\Gamma(E_m)/2)^2}
\end{equation}
whose energy-dependent width is given by the Brown-Rho parametrization
\cite{Brown81}
\begin{equation}
\label{eq:BR}
\Gamma(E_m) = \frac{ a (E_m - E_F)^2 }{(E_m-E_F)^2+b}
\end{equation}
where $E_F$ is the Fermi energy and $a$,$b$ are constants.
For the present purposes it suffices to use the parameters
$a=24$ MeV and $b=500$ MeV$^2$ originally proposed by 
Brown and Rho \cite{Brown81} with
$E_F = 15.96$ MeV and $E_{1s} = 39$ MeV for $^{12}$C.
Lapik\'as \etal obtained slightly different parameters for $b$ and
$E_{1s}$ by fitting the Tokyo data, but the effect upon $S_{1s}$
is insignificant.
Thus, one would expect approximately $11\%$ of the $1s$ strength
to be broadly distributed above 80 MeV missing energy, which is
consistent with the depletion factor attributed to short-range
correlations.
Therefore, in the absence of background, the $1s$ spectroscopic factor
deduced from a fitted normalization factor would be
\begin{equation}
\label{eq:1s}
S_{1s} = 2 N_{1s}(E_\text{min},E_\text{max})
\frac{ \int_0^\infty L_{1s}(E_m) dE_m } 
{ \int_{E_\text{min}}^{E_\text{max}} L_{1s}(E_m) dE_m }
\end{equation}

\begin{table}[hp]
\caption{Summary of data for $^{12}$C$(e,e^\prime p)$. 
$E_0$ is the beam energy and $T_p$ is the central proton kinetic energy. 
Quasiperpendicular kinematics with constant $(\omega,\bm{q})$ were used
unless a $Q^2$ range is given.
\label{table:data}}
\begin{ruledtabular}
\begin{tabular}{lrrrrrl}
data set & $E_0$ & $\Delta E_m(1p)$ & $\Delta E_m(1s)$ & $T_p$ & $Q^2$ & notes
\\ 
         & MeV   & MeV           & MeV  & MeV & (GeV/$c$)$^2$ & \\ \hline
Tokyo \cite{Nakamura76}       & 700     &  6-30 & 21-66 & 159 & 0.29 & \\
                              & 700     &       & 21-66 & 136 &      & \\
Saclay \cite{Mougey76}        & 497     & 15-22 & 30-50 &  87 & 0.16 & \\
Saclay \cite{Bernheim82}      & 500     & 15-22 &  NA   &  99 & 0.18 & constant $(\omega,\bm{q})$ \\
Saclay \cite{Bernheim82}      & 500     & 15-22 &  NA   &  99 & 0.09-0.32 & parallel; $1s$ not available \\
NIKHEF \cite{vdSteenhoven88a} & 285-481 & 15-22 & 30-39 &  70 & 0.02-0.26 & parallel \\
SLAC \cite{Makins94}          & 2015    & 15-25 & 30-80 & 600 & 1.11 & $1s$ not radiatively unfolded \\
JLab \cite{Dutta03}           & 2445    & 15-25 & 30-50 & 350 & 0.64 & inclusive bin with $10 \leq E_m \leq 80$ MeV available \\
                              & 2445    &       &       & 700 & 1.28 & \\
                              & 3245    &       &       & 970 & 1.84 &
\end{tabular}
\end{ruledtabular}
\end{table}


\section{Effective momentum approximation}
\label{sec:EMA}

\begin{figure}
\centering
\includegraphics[width=3.0in]{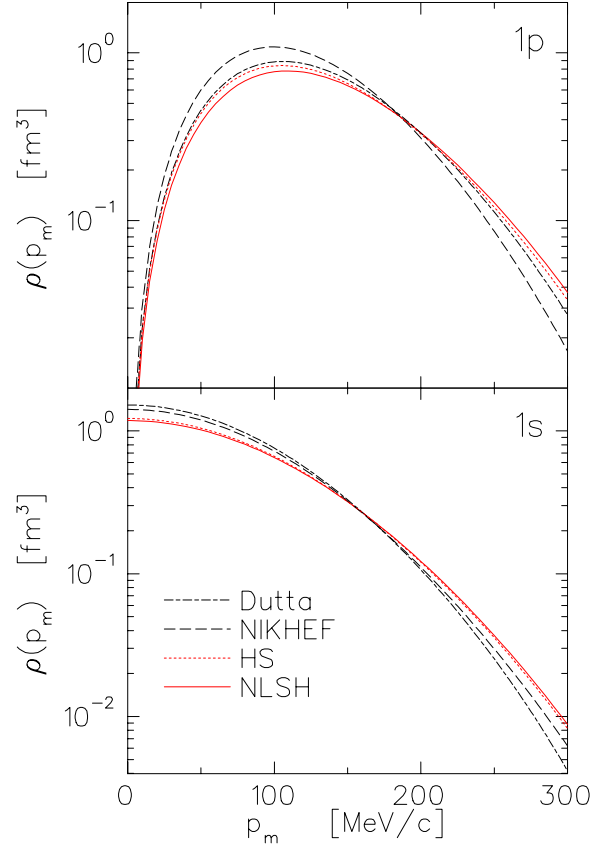}
\caption{(Color online) Momentum distribution for $^{12}$C$(e,e^\prime p)$.
}
\label{fig:12Cpm}
\end{figure}

Figure \ref{fig:12Cpm} compares momentum distributions for the relativistic
HS and NLSH models with the Woods-Saxon models used by  
Lapik\'as \etal \cite{Lapikas00} and by Dutta \etal \cite{Dutta03}.
The latter were based upon a NRDWIA analysis of the Saclay data for 
$T_p = 99$ MeV made by Bernheim \etal \cite{Bernheim82}.
The differences between the HS and NLSH wave functions are relatively small,
but the NIKHEF $1p$ wave function has a much stronger peak and declines
more rapidly with respect to $p_m$.
Similarly, the NIKHEF $1s$ wave function starts higher and falls faster
than the relativistic wave functions.
Interestingly, the Saclay wave functions show similar behavior but the effects,
compared with the NIKHEF wave functions, are smaller for the $1p$ and larger 
for the $1s$ wave function.
The large spread among these momentum distributions will obviously be
reflected in the corresponding DWIA calculations.

To evaluate the effect of variations of $\rho(p_m)$ upon DWIA calculations 
using both relativistic and nonrelativistic wave functions, it is simplest
to employ the EMA-noSV approach.
Several previous papers have shown that the differences between EMA and full
RDWIA calculations are relatively small for cross sections.
Although that approximation does not provide the best description of 
the left-right asymmetry, $A_{LT}$,
it has the advantage that the nonrelativistic wave function can be used
directly.

\begin{figure}
\centering
\includegraphics[angle=90,width=5.0in]{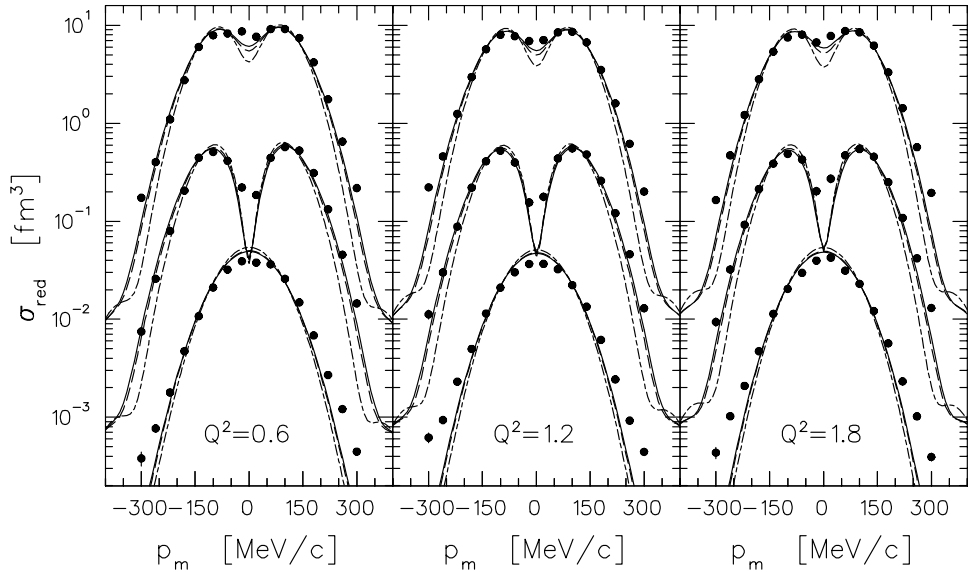}
\caption{EMA-noSV calculations for $^{12}$C$(e,e^\prime p)$ are compared
with quasiperpendicular data for $0.6 < Q^2 < 1.8$ (GeV/$c$)$^2$ from JLab 
\protect{\cite{Dutta03}}.
In each panel the top set of curves shows the inclusive 
$10 \leq E_m \leq 80$ MeV bin scaled by a factor of 20, 
the middle set shows the $p$-shell bin with $15 \leq E_m \leq 25$ MeV,
and the bottom set shows the $s$-shell bin with $30 \leq E_m \leq 50$ MeV
scaled by a factor of 0.1.
Solid, dashed, and dash-dotted curves are based upon NLSH, HS, and
NIKHEF wave functions, respectively.
}
\label{fig:Dutta-EMAnoSV}
\end{figure}

Figure \ref{fig:Dutta-EMAnoSV} compares EMA calculations using
NLSH, HS, and NIKHEF wave functions with the JLab data for $Q^2 = 0.6$,
1.2, and 1.8 (GeV/$c$)$^2$.
For each kinematics the top set are semi-inclusive data for $E_m < 80$ MeV, 
the second set shows a bin $15 \leq E_m \leq 25$ MeV dominated by the $1p$ 
shell, and the bottom set shows a bin $30 \leq E_m \leq 50$ MeV 
dominated by the $1s$ shell.
The curves were fit to data in the range $|p_m| \leq 200$ MeV/$c$ 
neglecting, at present, the mixing between the shells.
We will later find that filling of the $1p$ minimum and broadening of the 
$1s$ distribution can be described by mixing. 
These calculations employ the EDAD1 optical potential; other choices
of potentials from Cooper \etal \cite{Cooper93} produce small variations 
in the normalization factors with little dependence upon the choice of
overlap function and practically no discernible effect upon the curves.
The results for the two relativistic wave functions, NLSH and HS, are very 
similar but the fitted $p_m$ distributions for the NIKHEF wave function
tend to be too narrow, especially for the $1p$ shell.
Consequently, the $1p$ normalization factors listed in 
Table \ref{table:normperp_EMAnoSV} 
are systematically smaller for the NIKHEF wave function than for the NLSH or
HS wave functions. 
These normalization factors are correlated well with the peak of the momentum 
distribution, with the NLSH results largest and NIKHEF results smallest.

\begin{figure}
\centering
\includegraphics[width=3.0in]{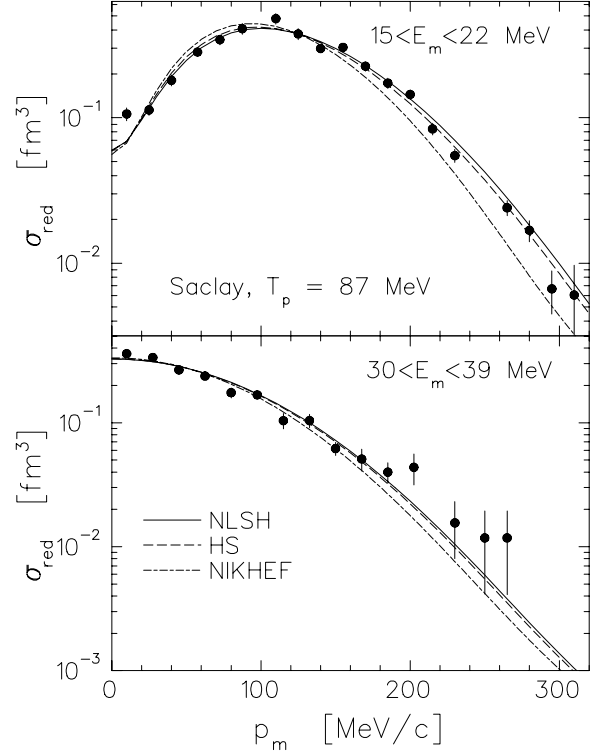}
\caption{EMA-noSV calculations for $^{12}$C$(e,e^\prime p)$ are compared
with quasiperpendicular data for $T_p = 87$ MeV from Saclay 
\protect{\cite{Mougey76}}.
Solid, dashed, and dash-dotted curves are based upon NLSH, HS, and
NIKHEF wave functions, respectively.
}
\label{EMA:Saclay_kin2}
\end{figure}

\begin{figure}
\centering
\includegraphics[width=3.0in]{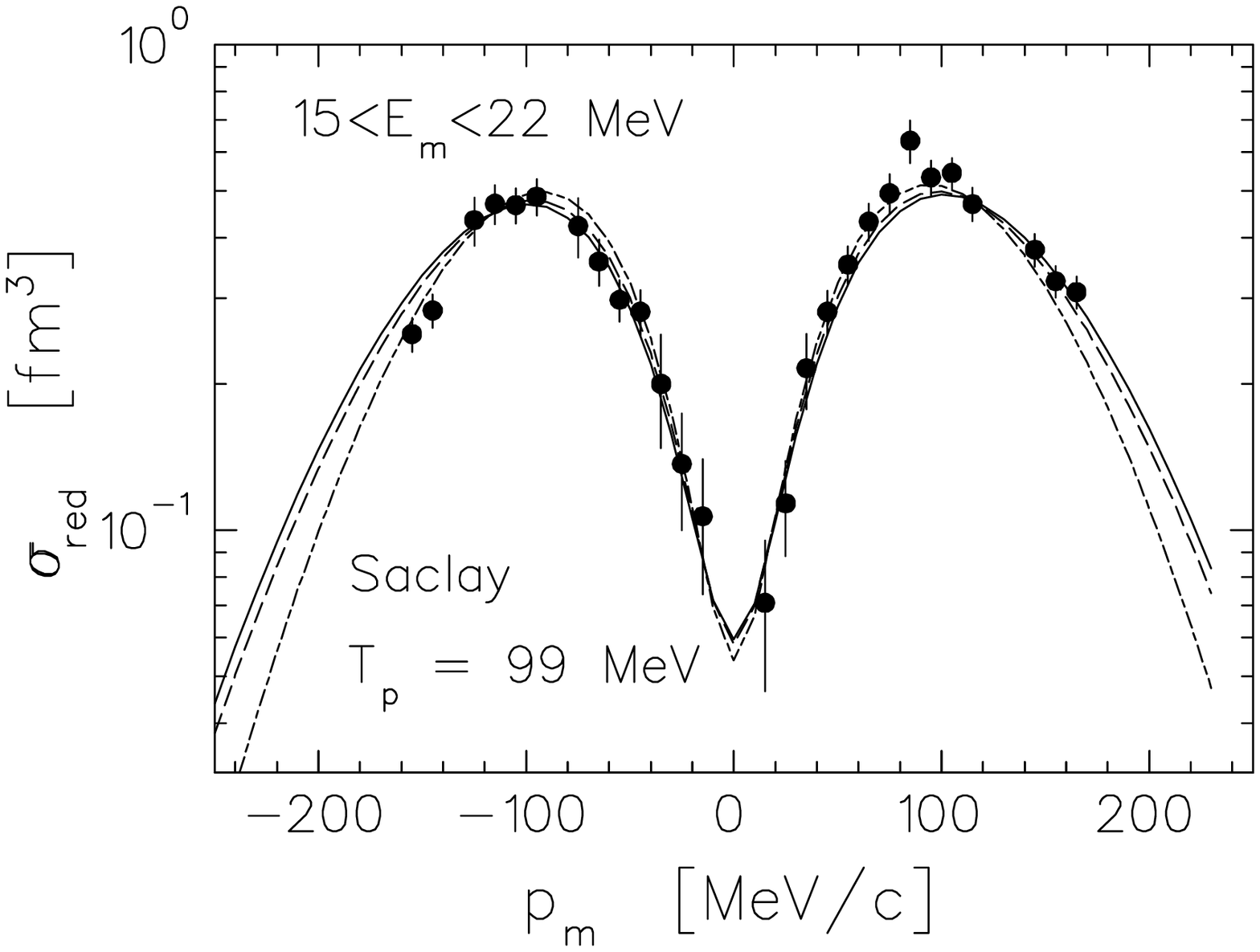}
\caption{EMA-noSV calculations for $^{12}$C$(e,e^\prime p)$ are compared
with quasiperpendicular data for $T_p = 99$ MeV from Saclay 
\protect{\cite{Mougey76}}.
Solid, dashed, and dash-dotted curves are based upon NLSH, HS, and
NIKHEF wave functions, respectively.
}
\label{EMA:Saclay_kin1}
\end{figure}

\begin{figure}
\centering
\includegraphics[width=2.5in]{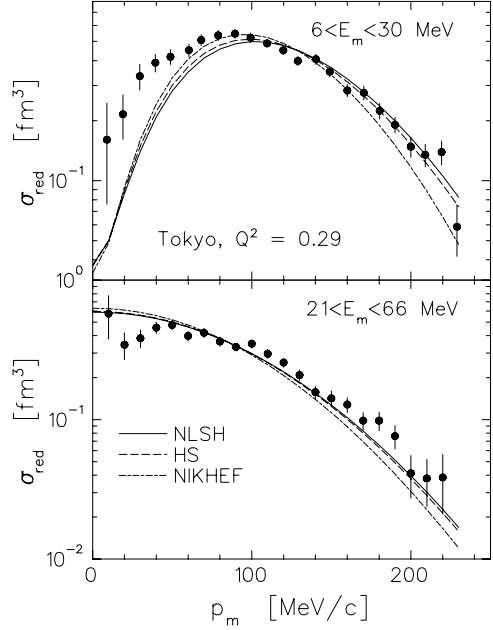}
\caption{EMA-noSV calculations for $^{12}$C$(e,e^\prime p)$ are compared
with quasiperpendicular data for $Q^2 = 0.29$ (GeV/$c$)$^2$ from Tokyo
\protect{\cite{Nakamura76}}.
}
\label{EMA:Tokyo}
\end{figure}

\begin{figure}
\centering
\includegraphics[width=2.5in]{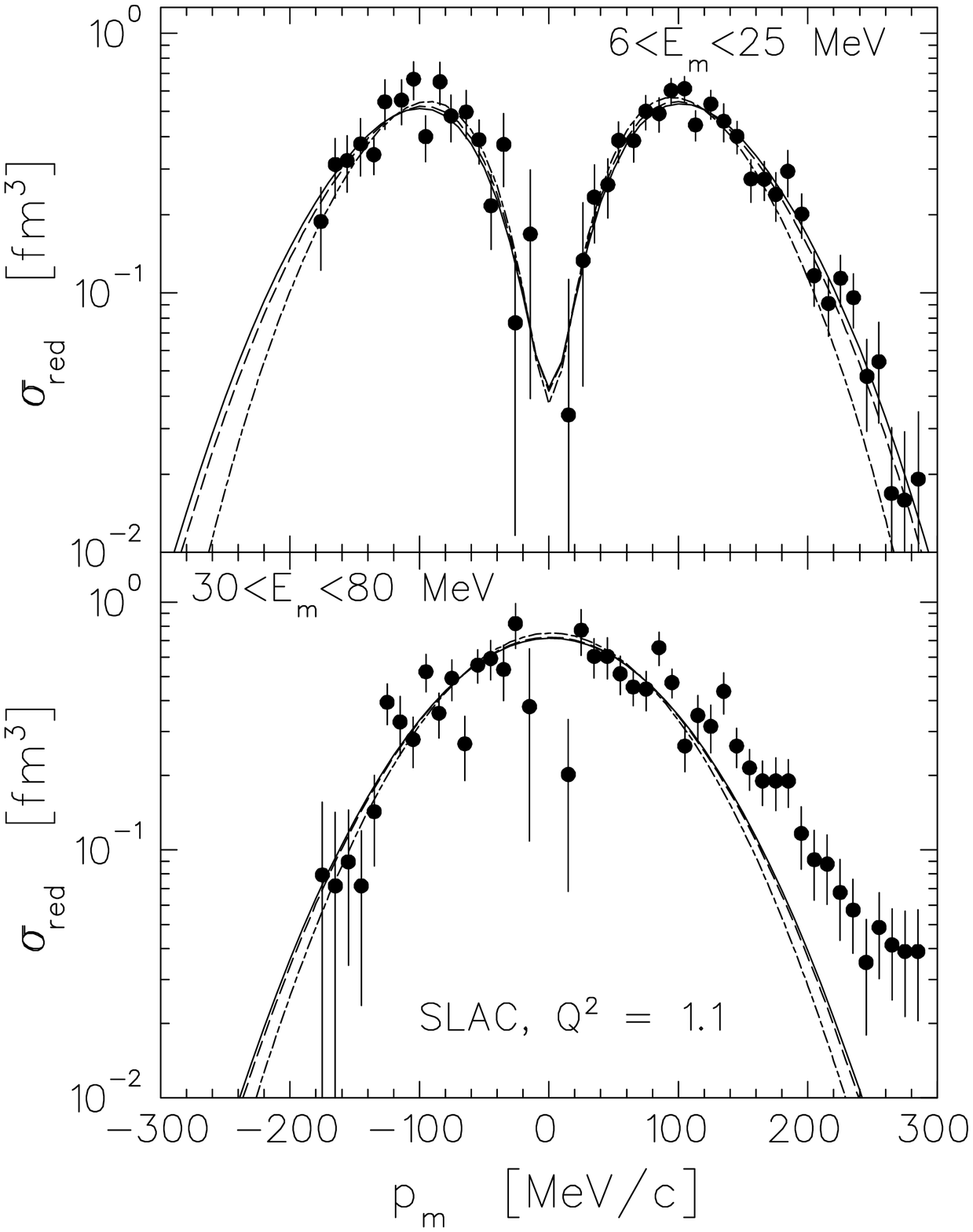}
\caption{EMA-noSV calculations for $^{12}$C$(e,e^\prime p)$ are compared
with quasiperpendicular data for $Q^2 = 1.1$ (GeV/$c$)$^2$ from SLAC
\protect{\cite{Makins94}}.
Solid, dashed, and dash-dotted curves are based upon NLSH, HS, and
NIKHEF wave functions, respectively.
}
\label{EMA:SLAC}
\end{figure}

Similar comparisons are shown in Figs. \ref{EMA:Saclay_kin2}-\ref{EMA:Tokyo} 
for several low $Q^2$ experiments performed using quasiperpendicular 
kinematics.
Except when noted otherwise, a normalization factor for each curve was 
obtained using a least-squares fit to the data for $|p_m| \leq 200$ MeV/$c$.
Fig. \ref{EMA:Saclay_kin2} shows that the relativistic wave functions 
describe the data for the Saclay experiment with $T_p = 87$ MeV in 
quasiperpendicular kinematics better than the NIKHEF wave functions,
which fall too rapidly with increasing $p_m$.
The results shown in Fig. \ref{EMA:Tokyo} for the Tokyo experiment with 
$Q^2 = 0.29$ (GeV/$c$)$^2$ are somewhat ambiguous: 
although the NLSH and HS wave functions give better fits to for
$p_m \gtrsim 60$ MeV/$c$, particularly for the ground-state, none of 
the calculations can describe the enhancement at lower $p_m$ seen for
the ground state or the dip for the $s$-state.
The low $p_m$ bulge for the ground-state data seems somewhat implausible 
and it is likely that there were difficulties in cleanly separating 
the $1p$ and $1s$ contributions to the low $p_m$ data.
Therefore, the $1p$ normalization for this set was fit to 
$50 \leq p_m \leq 200$ MeV/$c$.
Similar calculations are also compared with the SLAC data for $Q^2 = 1.1$
(GeV/$c$)$^2$ in Fig. \ref{EMA:SLAC}.
We again find that the $p_m$ distribution for the NIKHEF wave function is
too narrow to fit the $1p$ data well.
However, the $1s$ data remain well above all of the calculations for
large $p_m$.
Lapik\'as \etal \cite{Lapikas00} attributed this problem to $1p$ contamination
due to inadequate radiative unfolding.

The normalization factors obtained for single-component EMA-noSV fits 
are listed in Table \ref{table:normperp_EMAnoSV};
improved two-component RDWIA fits will be presented in the next section. 
The Tokyo result is omitted because the $1s$ admixture is appreciable.
We find that these factors are correlated well with the peak value
of the $\rho(p_m)$, such that the NIKHEF wave functions produce the
smallest and the NLSH wave functions the largest normalization factors.
Over a wide range of $Q^2$, including both early data at low $Q^2$ and
recent data for $Q^2 \sim 1$ (GeV/$c$)$^2$, the relativistic wave functions
fit the $p_m$ distributions relatively well while the NIKHEF wave
functions are systematically narrower than the data.
Thus, the data for quasiperpendicular kinematics consistently prefer the 
broader $p_m$ distributions and larger normalization factors obtained
with Dirac-Hartree wave functions.

\begin{table}
\caption{Normalization factors for $^{12}$C$(e,e^\prime p)(1p)^{-1}$ fit to 
quasiperpendicular data for the $1p$ bin using several overlap functions in
the EMAnoSV approximation.
Possible $1s$ contributions were neglected.
The first set uses the EDAD1 and the second the EDAIC optical potential.
\label{table:normperp_EMAnoSV}}
\begin{ruledtabular}
\begin{tabular}{lll|lll|lll}
           & $Q^2$          & $E_m$ & \multicolumn{3}{c|}{EDAD1} & \multicolumn{3}{c}{EDAIC}  \\
experiment & (GeV/$c$)$^2$  & MeV 
                           & NLSH  & HS    & NIKHEF & NLSH  & HS    & NIKHEF \\ \hline
Saclay     & 0.16   & 15-22 & 0.637 & 0.599 & 0.483  & 0.572 & 0.542 & 0.444 \\
Saclay     & 0.18   & 15-22 & 0.757 & 0.705 & 0.554  & 0.684 & 0.640 & 0.510 \\
JLab       & 0.6    & 15-25 & 0.934 & 0.878 & 0.697  & 0.909 & 0.856 & 0.682 \\
SLAC       & 1.1    & 15-25 & 0.852 & 0.791 & 0.613  & 0.887 & 0.822 & 0.633 \\
JLab       & 1.2    & 15-25 & 0.904 & 0.849 & 0.671  & 0.950 & 0.890 & 0.699 \\
JLab       & 1.8    & 15-25 & 0.901 & 0.854 & 0.689  & 0.886 & 0.841 & 0.679 \\
\end{tabular}
\end{ruledtabular}
\end{table}

\begin{figure}
\centering
\includegraphics[width=3.0in]{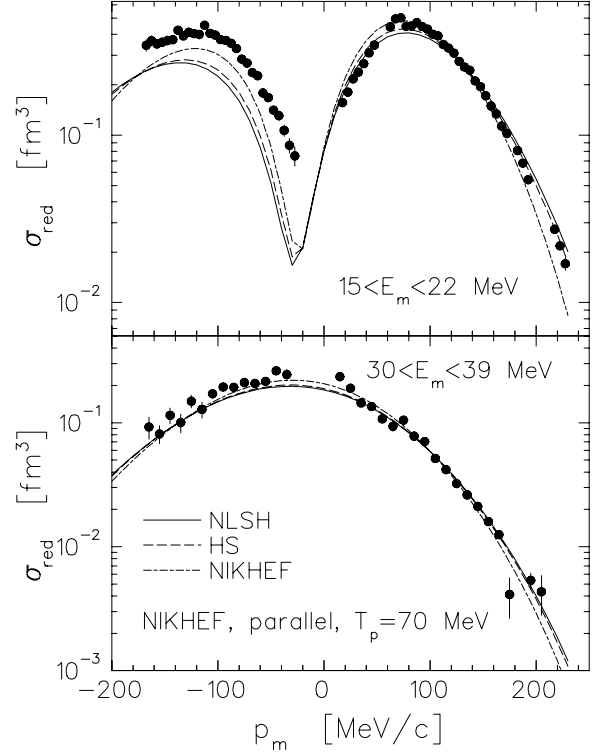}
\caption{EMA-noSV calculations for $^{12}$C$(e,e^\prime p)$ are compared
with data from from NIKHEF \protect{\cite{vdSteenhoven88a}} using parallel 
kinematics for $T_p = 70$ MeV/$c$.
}
\label{fig:NIKHEF-para}
\end{figure}

\begin{figure}
\centering
\includegraphics[width=3.0in]{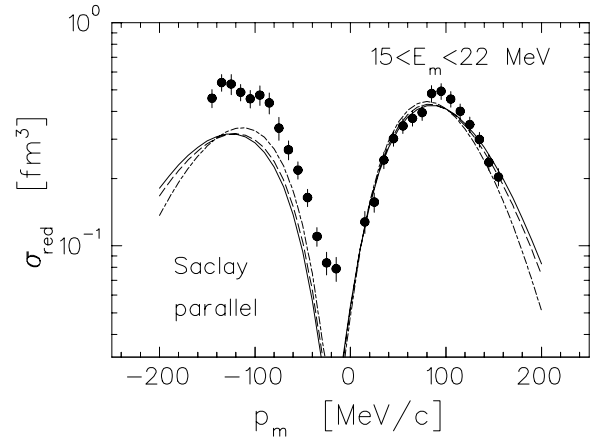}
\caption{EMA-noSV calculations for $^{12}$C$(e,e^\prime p)$ are compared
with data from from Saclay \protect{\cite{Bernheim82}} using parallel 
kinematics for $T_p = 99$ MeV.
Solid, dashed, and dash-dotted curves are based upon NLSH, HS, and
NIKHEF wave functions, respectively.
}
\label{fig:Saclay-para}
\end{figure}

The data for parallel kinematics, on the other hand, present special problems.
We omit the Mainz data \cite{Blomqvist95} because there appear to be 
significant normalization problems for that experiment \cite{Lapikas00}.
The NIKHEF and Saclay data for parallel kinematics are compared with
EMA-noSV calculations for NLSH, HS, and NIKHEF wave functions in
Figs. \ref{fig:NIKHEF-para} and \ref{fig:Saclay-para}.
The results for both experiments are similar but none of these calculations
reproduce the data for the $1p$ shell.
Due to obvious the difficulties for $p_m < 0$, the normalization factors
listed in Table \ref{table:normpara_EMAnoSV} were fit to 
$0 < p_m < 200$ MeV/$c$.
These factors are also correlated with the peak values of $\rho(p_m)$
but for the relativistic wave functions are somewhat lower than were
obtained using quasiperpendicular kinematics.
On the other hand, these fits underestimate the cross sections for
$p_m <0$ and larger $Q^2$.

\begin{table}
\caption{Normalization factors for $^{12}$C$(e,e^\prime p)(1p)^{-1}$ fit to 
data for parallel kinematics using several overlap functions in the
EMAnoSV approximation.
These results were extracted for the $15 \leq E_m \leq 22$ MeV bin
using NLSH wave functions and the EDAD1 optical potential.
\label{table:normpara_EMAnoSV}}
\begin{ruledtabular}
\begin{tabular}{|l|l|l|l|l|}
experiment &  $T_p$ & NLSH & HS & NIKHEF  \\ \hline
NIKHEF & 70 & 0.635 & 0.616 & 0.541  \\
Saclay & 99 & 0.685 & 0.636 & 0.505  \\
\end{tabular}
\end{ruledtabular}
\end{table}

\begin{figure}[p]
\centering
\includegraphics[width=3.0in]{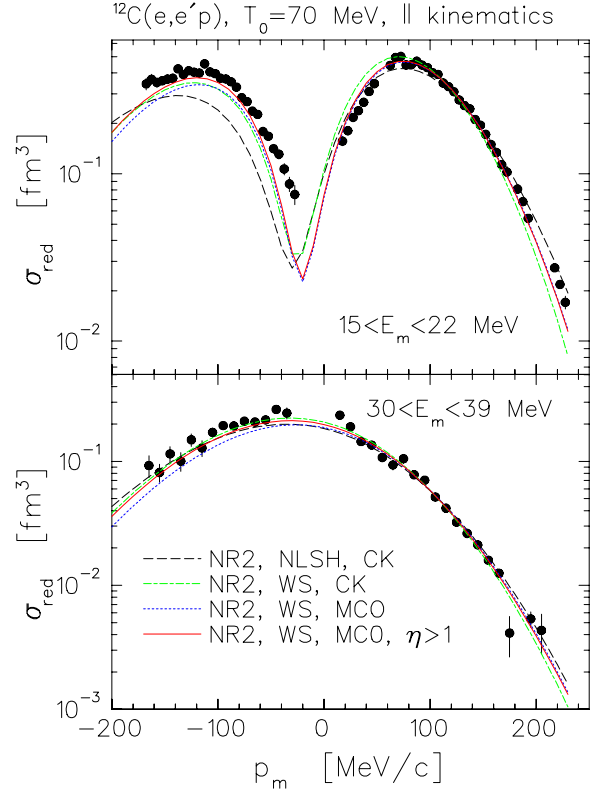}
\caption{(Color online) 
NR2 calculations for $^{12}$C$(e,e^\prime p)$ are compared
with data from from NIKHEF \protect{\cite{vdSteenhoven88a}} using parallel 
kinematics for $T_p = 70$ MeV/$c$.
WS refers to the NIKHEF Woods-Saxon wave function and $\eta>0$ shows the
effect of enhancing transverse components of the current opertor.
}
\label{fig:NR2-para}
\end{figure}

In order to improve standard NRDWIA calculations for parallel kinematics, 
van der Steenhoven \etal \cite{vdSteenhoven88a,Ireland94} resorted to 
adjusting the overlap function, the optical model, and the current operator 
simultaneously.
The NIKHEF analysis used the nonrelativistic expansion of the current
operator in powers of $p/m$ proposed by McVoy and van Hove \cite{McVoy};
we verified that at second order this expansion, denoted NR2, gives results 
that are practically indistinguishable from our EMA-noSV calculations.
Their analysis also used a nonrelativistic optical potential produced
by Comfort and Karp \cite{Comfort80}, denoted CK, and a 
Perey factor \cite{Perey62} for the ejectile with $\beta=0.85$ fm. 
Figure \ref{fig:NR2-para} compares several variations of the NR2 calculations 
that are similar to those of van der Steenhoven {\it et al}.
The dashed curve uses this optical potential and current operator with
the upper component of the NLSH wave function and fails to reproduce 
the data for parallel kinematics with $p_m <0$; the results are similar
to those shown in Fig. \ref{fig:NIKHEF-para} in the EMA approach.
The narrower NIKHEF momentum distribution improves the fit to the data but
appears to be shifted to the left ---
a shift to the right was accomplished by modifying the optical potential.
This modified potential, denoted MCO in Ref. \cite{vdSteenhoven88a}, was 
intended to simulate channel coupling in the final state but our more
detailed analysis of channel coupling \cite{Kelly99a} gave much smaller 
effects for this reaction; hence, we regard this modification of the
optical potential to be somewhat {\it ad hoc}.
Finally, van der Steenhoven \etal inserted an enhancement factor for
transverse components of the current operator and adjusted that factor to
fit the peak for negative $p_m$.
Applying this factor without further adjustment does provide a reasonably 
accurate fit to the reduced cross section for parallel kinematics.

We have not attempted to tune this fit because we doubt that it is really
possible to distinguish between variations of the overlap function, 
the current operator, and the final-state interactions using these data.
Nor does it appear possible to fit the low $Q^2$ data for both parallel 
and quasiperpendicular kinematics simultaneously using unique overlap 
functions and global optical potentials fit to proton elastic scattering.
The relativistic wave functions provide goods fits to the $p_m$ dependence
of the data for quasiperpendicular kinematics even if the normalization 
factors for $Q^2 \ge 0.6$ (GeV/$c$)$^2$ are significantly larger than those 
for low $Q^2$.
On the other hand, the normalization factors for the two Saclay
experiments with quasiperpendicular kinematics do not agree well
with each other either.
Therefore, the failure of EMA calculations with NLSH wave functions to 
reproduce the $p_m$ dependence of data for parallel kinemtics with 
$T_p \lesssim 100$ MeV seems to be a more serious problem than the variation
of normalization factors with $Q^2$.
Recognizing that off-shell ambiguities are more serious for parallel 
kinematics because $x=Q^2/2m \omega$ varies over a wide range and 
expecting the reaction mechanism to be more reliable for large $Q^2$,
we believe that the reaction mechanism for parallel kinematics with 
$T_p \lesssim 100$ MeV is not sufficiently reliable to justify fitting 
the bound-state wave function and that it is likely that in such analyses
variations of the overlap function would tend to compensate for
deficiencies in the reaction model.

If one wants to fit the momentum distribution for a discrete single-hole 
state to $(e,e^\prime p)$ data, it is probably best to use quasifree 
kinematics where $(\omega,\bm{q})$ is constant and $Q^2$ is as large as 
possible.
The reliability of the reaction mechanism should improve as $Q^2$ increases 
and the use of quasifree kinematics should minimize distortion of 
$\rho(p_m)$ due to variation of current operator off shell.
It would then be of interest to test the reaction mechanism by comparison
with data for parallel kinematics with large $T_p$.
Ideally such tests should have sufficient resolution in $E_m$ to isolate
discrete states without the complications of incoherent mixtures. 
Unfortunately, no high-resolution $(e,e^\prime p)$ data for parallel 
kinematics with $T_p > 200$ MeV and sufficient coverage in $p_m$ are available.

\section{RDWIA analysis}
\label{sec:RDWIA}

\subsection{Normalization factors}
\label{sec:norm}

\subsubsection{JLab}
\label{sec:norm-JLab}

\begin{figure}
\centering
\includegraphics[angle=90,width=5in]{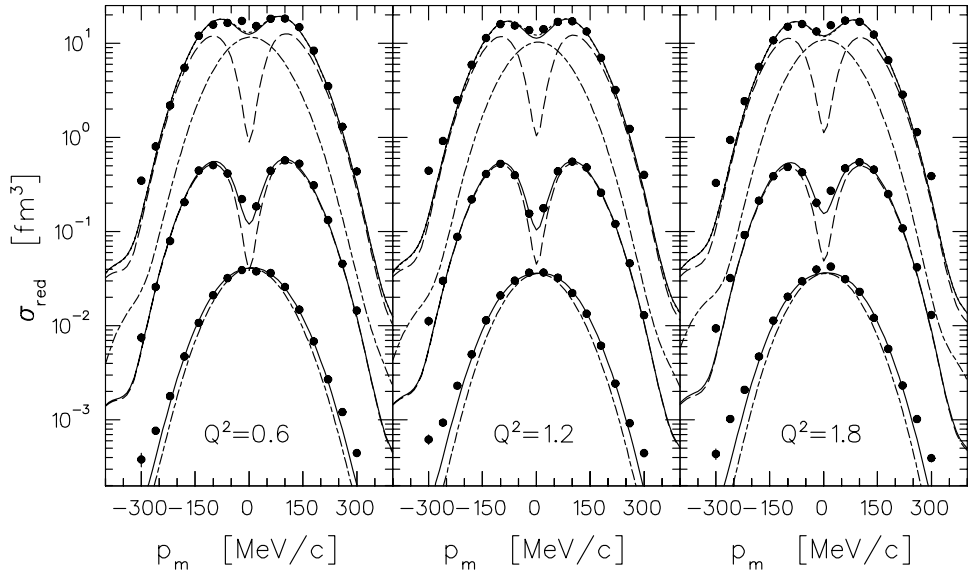}
\caption{
Contributions of $1p_{3/2}$ and $1s_{1/2}$ fitted to JLab data for
$^{12}$C$(e,e^\prime p)$.  
In each panel the top set of curves shows the inclusive 
$10 \leq E_m \leq 80$ MeV bin scaled by a factor of 20, 
the middle set shows the $p$-shell bin with $15 \leq E_m \leq 25$ MeV,
and the bottom set shows the $s$-shell bin with $30 \leq E_m \leq 50$ MeV
scaled by a factor of 0.1.
The solid curves show the total fit, the dashed curves show the $1p$
contributions to the inclusive and $p$-shell bins, and the dash-dotted 
curves show the $1s$ contributions to the inclusive and $s$-shell bins.
In addition, scale-factor fits for the IPSM  are shown as dotted curves
for the inclusive bin but are practically indistinguishable from the
solid curves.
All calculations use NLSH wave functions and EDAD1 optical potentials
in full RDWIA.}
\label{fig:Dutta_nlsh}
\end{figure}

Full RDWIA fits to the JLab data \cite{Dutta03} are shown in 
Fig. \ref{fig:Dutta_nlsh} using NLSH wave functions, EDAD1 optical potentials,
and the CC2 current operator in Coulomb gauge.
For the inclusive bin, these fits are slightly better than the corresponding
EMA-noSV fits shown in Fig. \ref{fig:Dutta-EMAnoSV} because full RDWIA
calculations reproduce the left-right asymmetry better than the EMA-noSV
approximation \cite{Kelly05a}.
These fits for the $1p$ and $1s$ bins are significantly better than 
single-component fits because neither bin represents a pure single-particle
configuration.
The $1s$ contribution to the $1p$ bin fills in the minimum and rounds the
peaks. 
Similarly the $1p$ contribution to the $1s$ bin broadens the momentum
distribution --- the reduced $1s$ component provides a better fit for small 
$p_m$ while the $1p$ component improves the fit to larger $p_m$.
However, for $p_m > 200$ MeV/$c$ the data for both the $s$-shell and inclusive
bins remain significantly above these two-component fits.
Similar data for $^{16}$O$(e,e^\prime p)$ \cite{Liyanage01,Fissum04} covering
a broad range of missing energy show that for $p_m \gtrsim 150$ MeV/$c$
there is much more strength at large $E_m$ than can be attributed to
single-nucleon knockout from the $s$-shell.
The continuum fraction increases with $p_m$, which may be simulated in the 
two-component model by a $1p$ admixture.
Assuming that the $1s$ quasihole momentum distribution is similar
to the Dirac-Hartree wave function, it would appear that the continuum
contribution to $E_m > 30$ MeV has a broader momentum distribution than
the $1p_{3/2}$ wave function used in these two-component fits.
Although part of this may be contributed by $1p$ fragmentation, much of the
fitted $1p$ strength is probably a surrogate for multinucleon continuum.
Recent SCGF estimates suggest that approximately $68\%$ of the 
correlated continuum is in partial waves with $\ell \leq 1$ \cite{Muther04}.
No doubt higher partial waves begin to contribute to the continuum as $p_m$ 
increases and it is likely that the $\ell=1$ continuum is also
broader than the $1p_{3/2}$ quasihole momentum distribution.

\begin{figure}
\centering
\includegraphics[angle=90,width=5in]{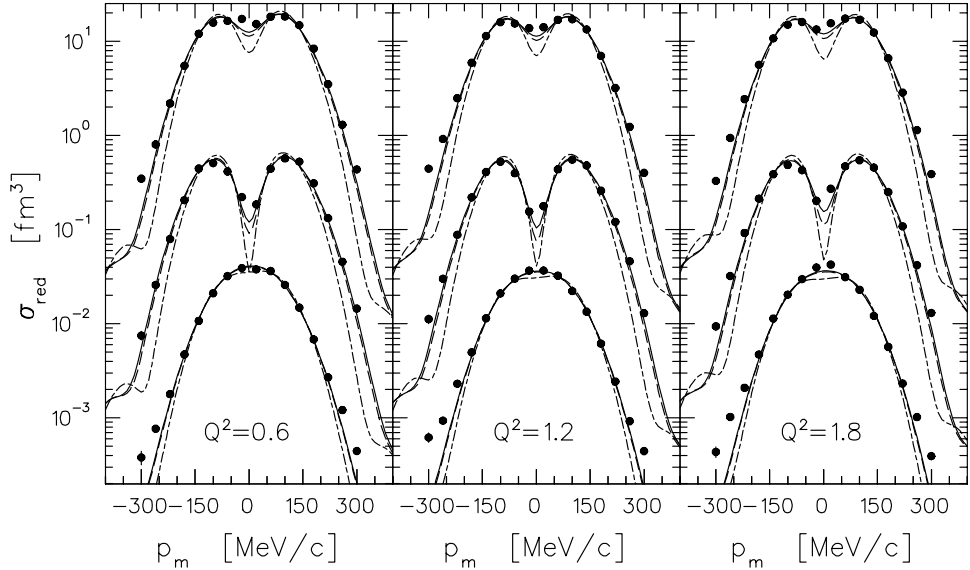}
\caption{RDWIA fits to the JLab data for JLab data for $^{12}$C$(e,e^\prime p)$
are compared for NLSH (solid), HS (dashed), and NIKHEF (dash-dot) wave functions. 
In each panel the top set of curves shows the inclusive 
$10 \leq E_m \leq 80$ MeV bin scaled by a factor of 20, 
the middle set shows the $p$-shell bin with $15 \leq E_m \leq 25$ MeV,
and the bottom set shows the $s$-shell bin with $30 \leq E_m \leq 50$ MeV
scaled by a factor of 0.1.}
\label{fig:Dutta_fits}
\end{figure}

RDWIA fits to the JLab data using NLSH, HS, or NIKHEF wave functions are 
compared in Fig. \ref{fig:Dutta_fits}.
The two Dirac-Hartree wave functions provide very similar fits with a
slight preference for NLSH.
Mixing between $1p$ and $1s$ contributions improves fits to the
$1s$ bin using NIKHEF wave functions by flattening the small $p_m$
and lifting the large $p_m$ regions.
However, two-component fits using NIKHEF wave functions were impossible 
for the $1p$ bin because the $1s$ normalizations were negative; 
hence, the $1p$ NIKHEF fits shown here simplify scale RDWIA calculations 
for $1p_{3/2}$.
Furthermore, the $1p$ NIKHEF fits are even worse and their normalization 
factors are even smaller in RDWIA than in EMA-noSV: 
with EDAD1 those normalization factors are 0.66, 0.63, and 0.66 for RDWIA
compared with 0.70, 0.67, and 0.69 for EMA-noSV at the three $Q^2$ settings. 
This reduction occurs because the RDWIA calculations for inherently
nonrelativistic Woods-Saxon wave functions were made by replacing 
the Perey factor with the Darwin factor obtained using the spin-orbit
potential; this procedure is explained in Ref. \cite{Kelly05a}.
However, because the spin-orbit potential used by van der Steenhoven
\etal \cite{vdSteenhoven88a} is stronger than that for NLSH or HS 
Dirac-Hartree wave functions, the smaller Darwin factor leads to a 
narrower momentum distribution for RDWIA than for EMA-noSV and, 
consequently, smaller normalization factors.
The fits for the $1p$ bin overshoot the peaks and then fall too
rapidly as $p_m$ increases and these problems carry over, in slightly
diluted form, to the inclusive bin.
Therefore, we reject those parametrizations of the $1p_{3/2}$ and 
$1s_{1/2}$ overlap functions.

RDWIA normalization factors fit to the JLab reduced cross section data for 
$^{12}$C$(e,e^\prime p)$ in three bins of missing energy are collected in 
Table \ref{table:Dutta_norm}.
These results were obtained using EDAD1 optical potentials but
the corresponding results for EDAD2, EDAD3, EDAIC, or EEI optical potentials
differ by less than $\pm 5\%$ with systematics similar to 
Table \ref{table:normperp_EMAnoSV}.
The differences between NLSH and HS results are small for $1p_{3/2}$ and
modest for $1s_{1/2}$.
Most of the discussion will be based upon the NLSH results because they
provide slightly better fits to the data.

\begin{table}
\caption{Normalization factors for $^{12}$C$(e,e^\prime p)$ fit to JLab data
for three bins of missing energy and three values of $Q^2$ in (GeV/$c$)$^2$.
The columns labeled {\it ave} is averaged with respect to $Q^2$.
The first set uses NLSH and the second HS wave functions.
Both sets use EDAD1 optical potentials.  
\label{table:Dutta_norm}}
\begin{ruledtabular}
\begin{tabular}{llllll|llll}
& & \multicolumn{4}{c|}{NLSH} & \multicolumn{4}{c}{HS} \\ 
state &  $E_m$ bin & $Q^2 = 0.6$ & $Q^2 = 1.2$ & $Q^2 = 1.8$ & ave 
                   & $Q^2 = 0.6$ & $Q^2 = 1.2$ & $Q^2 = 1.8$ & ave  \\ \hline
$1p_{3/2}$ & 15 -- 25 MeV & 0.885 & 0.868 & 0.847 & 0.87  
                          & 0.852 & 0.836 & 0.830 & 0.84 \\
           & 30 -- 50 MeV & 0.060 & 0.078 & 0.077 & 0.072 
                          & 0.063 & 0.058 & 0.062 & 0.061 \\
           & 10 -- 80 MeV & 1.021 & 1.001 & 0.977 & 1.00 
                          & 1.008 & 0.985 & 0.973 & 0.99 \\ \hline
$1s_{1/2}$ & 15 -- 25 MeV & 0.133 & 0.102 & 0.198 & 0.14  
                          & 0.082 & 0.050 & 0.110 & 0.08 \\
           & 30 -- 50 MeV & 0.669 & 0.616 & 0.667 & 0.65  
                          & 0.628 & 0.575 & 0.620 & 0.61 \\
           & 10 -- 80 MeV & 0.953 & 0.887 & 1.009 & 0.95  
                          & 0.818 & 0.757 & 0.845 & 0.81 \\ \hline
IPSM       & 10 -- 80 MeV & 0.981 & 0.991 & 0.984 & 0.985 
                          & 0.949 & 0.914 & 0.940 & 0.934
\end{tabular}
\end{ruledtabular}
\end{table}

A rather large fraction, approximately $87\%$, of the IPSM $1p$ strength 
is found in the lowest $E_m$ bin. 
If we assume that fragmentation of the $1p$-strength by collective modes
is largely confined to excitation energies below 10 MeV, 
the $87\%$ we find for the $1p$-shell is consistent with SCGF estimates.
The additional $1p$ strength fit to the reduced cross section for 
larger $E_m$ probably represents continuum contributions with $\ell > 0$ 
that should not be added to the valence $1p$ strength.

The $1s$ contribution to the $p$-shell bin is more than predicted by 
Eqs. (\ref{eq:L}-\ref{eq:BR}), but could arise either from resolution, 
discrete $\frac{1}{2}^+$ states, or a stronger low-$E_m$ tail.
Approximately $65\%$ of the $1s$ IPSM strength appears in the 
$30 < E_m < 50$ MeV bin and corresponds to a net occupancy of $94\%$ 
when scaled according to Eq. (\ref{eq:1s}) to account for the 
spreading width.
However, this is probably an overestimate of the $1s$ spectroscopic factor 
because one would expect approximately $10\%$ of the $1s$ strength to
lie beyond 80 MeV missing energy; hence, if we scale $N_{1s}$ for the 
NLSH fit to the inclusive bin we would obtain more than $100\%$.
Although the overestimation appears to be smaller for the HS fits, 
neither accounts for continuum contributions from rescattering or
two-body currents that could artificially enhance $N_{1s}$.
The difficulties in reproducing the cross sections for $p_m > 200$ MeV/$c$ 
in the $E_m > 30$ MeV region also suggest additional continuum 
contributions with broader momentum distribution that are not
described by these two-component fits.
Therefore, it is reasonable to assume that the depletion of the $1s$
shell by short-range correlations is similar to that for the $1p$ shell
and to attribute the additional strength in the inclusive cross section
to continuum processes that are not related directly to the single-hole
spectral function. 

The final line of Table \ref{table:Dutta_norm} lists IPSM scale factors 
fit to the $E_m < 80$ MeV data according to Eq. (\ref{eq:NIPSM}); 
these results are in reasonable agreement with the weighted average of the 
$1p$ and $1s$ normalization factors. 
The corresponding fits are shown in Fig. \ref{fig:Dutta_nlsh} as
dotted curves that are practically indistinguishable from the two-component
fits to the inclusive data.
Therefore, we find approximately $98.5\%$ of the IPSM strength in 
$E_m < 80$ MeV, which is considerably more than the $85\%$ one expects
after accounting for short-range correlations.
It is likely that much of the extra strength observed for 
$E_m > 30$ MeV arises from processes other than single-nucleon knockout 
from the $1s$-shell, such that the total single-hole strength for 
$E_m < 80$ MeV is overestimated by this analysis.

Under these conditions, we consider $N_{1p}(15,25)=0.87$ to be the 
most reliable estimate of the depletion of IPSM orbitals in 
$^{12}$C by short-range correlations.
This estimate is consistent with the direct measurements of correlated
continuum made by Rohe \etal \cite{Rohe04} using parallel kinematics and 
the analysis of those data by M\"uther and Sick \cite{Muther04} based upon 
SCGF.
The extra yield observed by Dutta \etal \cite{Dutta03} in
quasiperpendicular kinematics should then be attributed to
two-body currents and rescattering processes in which some of the flux 
absorbed by the optical potential is distributed to final states that 
remain within the experimental acceptance.
Several studies suggest that rescattering contributions to the continuum
are minimized for parallel kinematics \cite{Sick97,Rohe04,Barbieri04}.
These background contributions affect the interpretation of
nuclear transparency measurements, which will be considered in 
Sec. \ref{sec:transparency}.

\begin{figure}
\centering
\includegraphics[angle=90,width=5in]{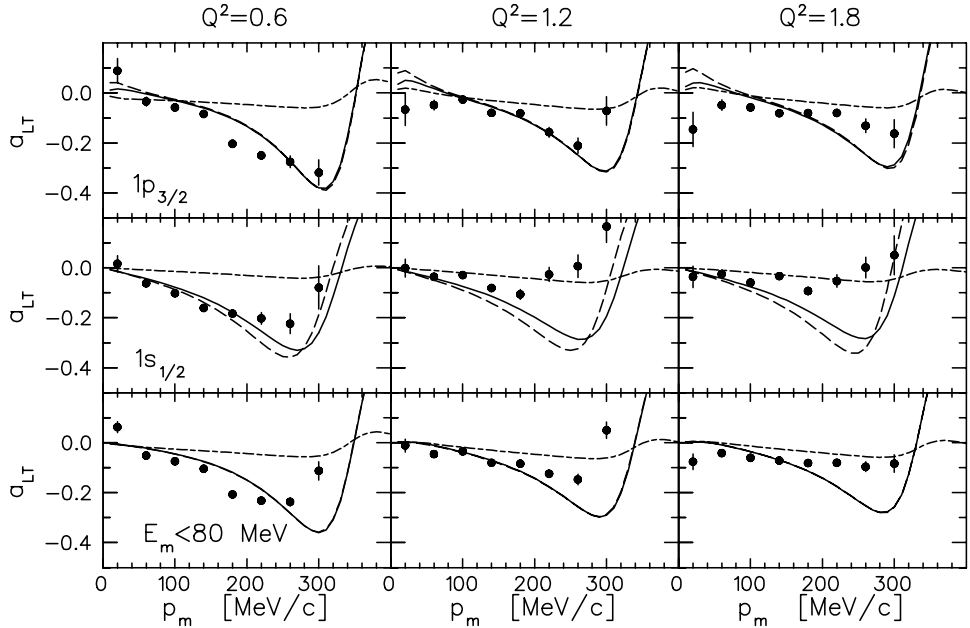}
\caption{RDWIA calculations for the left-right asymmetry in reduced
cross section for quasiperpendicular kinematics with
$0.6 \leq Q^2 \leq 1.8$ (GeV/$c$)$^2$.
NLSH wave functions, EDAD1 optical potentials, and the CC2 current
operator were used.
The dashed curves show single-component calculations while results
for two-component $1p+1s$ fits are shown as solid curves.
The dash-dotted curves show two-component fits in EMA-noSV.}
\label{fig:alt_fit}
\end{figure}

For the sake of completeness, we also show in Fig. \ref{fig:alt_fit} the
left-right asymmetry in reduced cross section 
\begin{equation}
\label{eq:alt}
a_{LT} = 
\frac{\sigma_\text{red}(\phi=0) - \sigma_\text{red}(\phi=\pi)}
     {\sigma_\text{red}(\phi=0) + \sigma_\text{red}(\phi=\pi)}
\end{equation}
where the azimuthal angle $\phi=0$ corresponds to an ejectile momentum 
in the electron scattering plane between the beam direction and the 
momentum transfer.
The sensitivity of this quantity to spinor distortion was studied 
in some detailed in Ref. \cite{Kelly05a} for pure single-particle
configurations.
Figure \ref{fig:alt_fit} shows that the $1p_{3/2}$ admixture for the
$s$-shell bin provides slightly better fits to the data for
$p_m \lesssim 300$ MeV/$c$.
That effect is small for the $p$-shell bin because the $1p$ admixture
is small and is also small for the inclusive bin because the 
asymmetries for the two components are similar, depending more upon
the Dirac potentials than upon the details of the bound-state
wave functions.
Two-component fits based upon EMA-noSV calculations demonstrate
that $a_{LT}\approx 0$ without spinor distortion because the 
characteristic left-right asymmetry for electron scattering by
a moving free proton is removed by using the reduced cross section.
The remaining asymmetry due to dynamical enhancement of lower
components of Dirac spinors is described very well by RDWIA 
calculations.
As argued in Ref. \cite{Kelly05a}, the flattening of the $p_m$
distribution for $a_{LT}$ in the $s$-shell and inclusive bins 
for increasing $Q^2$ is probably caused by continuum contributions
that need not share the characteristic asymmetry of single-nucleon
knockout.

\subsubsection{Previous experiments}
\label{sec:RDWIA-lowQ2}

The RDWIA analysis of the JLab data presented in the preceding section
provides a satisfying degree of internal consistency over a broad
range of $Q^2$.
Unfortunately, similar analyses for older data sets present a variety
of consistency problems that cannot be resolved by simply by choosing
the best overlap functions or optical potentials.
Therefore, in this section we consider those data in chronological
order using only a single set of options, namely NLSH and EDAD1, in
the RDWIA framework.

\begin{figure}
\centering
\includegraphics[width=2.5in]{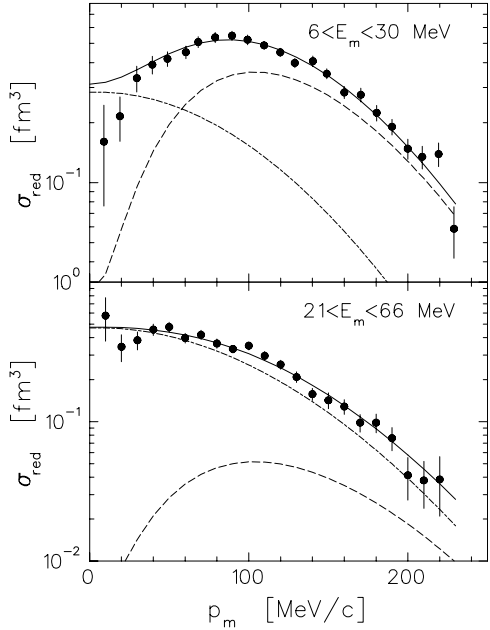}
\caption{RDWIA fit to Tokyo data \protect{\cite{Nakamura76}} using
NLSH wave functions.
Solid curves show the fits while dashed and dash-dotted curves show
$1p$ and $1s$ contributions.
}
\label{fit:Tokyo}
\end{figure}

The fits shown in Fig. \ref{fit:Tokyo} appear to describe the Tokyo 
data \cite{Nakamura76} very well, 
but the $1s$ contribution to the $p$-shell bin is implausibly large.
With NLSH wave functions and EDAD1 optical potential, we find
$(N_{1p},N_{1s}) = (0.56,0.42)$ for $6 \leq E_m \leq 30$ MeV and
$(N_{1p},N_{1s}) = (0.08,0.69)$ for $21 \leq E_m \leq 66$ MeV.
Apparently there was an experimental problem in defining the $p$-shell 
or there is an unidentified problem in the reaction mechanism that 
enhances the cross section for small $p_m$.
Because this is the only quasiperpendicular data set with severe
problems in the $p$-shell bin for small $p_m$, we believe that energy
resolution is a more likely explanation than reaction mechanism.
The single-component fit to $p_m > 50$ MeV shown in 
Fig. \ref{EMA:Tokyo} corresponds to $N_{1p} = 0.77$, which is
more consistent with the other quasiperpendicular data for this bin. 
The $1p$ contribution to the $1s$ bin, on the other hand, is fairly small.
Scaling the $1s$ fit for $21 \leq E_m \leq 66$ MeV according to
Eq. (\ref{eq:1s}) gives a net occupancy of $80\%$, which is reasonable
but smaller than the corresponding JLab result.
Perhaps the continuum contamination was smaller for the Tokyo experiment.

RDWIA provides no qualitative improvement over EMA-noSV fits to
NIKHEF or Saclay data for parallel kinematics.
As discussed before, we believe the severe variation of Bjorken $x$ in 
parallel kinematics distorts momentum distributions fit to $(e,e^\prime p)$
data for low $Q^2$ using familiar off-shell extrapolations of the 
single-nucleon current operator.
Therefore, we do not consider those data further. 
Conversely, single-component RDWIA fits to the Saclay data for 
quasiperpendicular kinematics are similar to those shown in Figs. 
\ref{EMA:Saclay_kin2}-\ref{EMA:Saclay_kin1}.
There is no evidence for $1s$ contamination of the 
$15 \leq E_m \leq 25$ MeV bin and the $30 \leq E_m \leq 50$ MeV bin 
is too narrow and the range of $p_m$ is too small to find significant 
$1p$ strength there.
Furthermore, mixed fits sometimes produce negative contamination factors 
with large uncertainties; 
hence, only single-component fits are considered for the 
Saclay quasiperpendicular data.
We find $1p$ normalization factors of 0.63 for Ref. \cite{Mougey76} or 
0.75 for Ref. \cite{Bernheim82} that are somewhat smaller than the 
JLab results but are not in especially good agreement with each other
either.
Similarly, $N_{1s}=0.46$ for 30-50 MeV corresponds to an occupancy
of $67\%$, which is also somewhat less than the JLab result.
Given that we believe the reaction mechanism is more reliable at 
higher $Q^2$ and that there are unresolved problems in reproducing
data in parallel kinematics at the same $Q^2$, we are not especially
alarmed by a 10-20\% discrepancy between $Q^2 < 0.2$ and $Q^2 > 0.6$
(GeV/$c$)$^2$.

Finally, the statistical quality of the $1p$ SLAC data for 
$Q^2 = 1.1$ (GeV/$c$)$^2$ do not support an extraction of a 
$1s$ admixture either.
Nor did we attempt to fit the $1p$ contribution to the 
$30 \leq E_m \leq 80$ MeV bin because the enhancement for large positive 
$p_m$ is probably an artifact of inadequate radiative unfolding.
Single-component fits using RDWIA calculations are very similar
to the EMA fits shown in Fig. \ref{EMA:SLAC} and the figures are omitted.
The SLAC $1p$ normalization factor for the $p$-shell bin is 0.86 for 
NLSH and EDAD1, which is perfectly consistent with the JLab results.
However, $N_{1s} = 1.2$ for 30-80 MeV is much larger than the JLab
result and scaled up to unrestricted $E_m$ would represent 140\%
of IPSM, suggesting a rather substantial underlying continuum.

\subsubsection{$Q^2$ dependence of spectroscopic factors}
\label{sec:Q2}

Lapik\'as \etal \cite{Lapikas00} proposed that the $Q^2$ variation 
seen in their analysis of the spectroscopic factors for $^{12}$C 
could be explained in terms of the resolution with which the 
electromagnetic knockout reaction probes the structure of a
quasiparticle, such that the true spectroscopic factor would be
observed in the limit $Q^2 \longrightarrow \infty$.
Although this model was expressed in terms of a scale-dependent
renormalization of the spectroscopic factor, it would be more
appropriate to attribute such an effect, if present, to the 
single-nucleon current operator instead of to the spectroscopic
factor.
The overlap function for single-nucleon knockout takes the form
\begin{equation}
\langle \Psi^{(A-1)}_\alpha | \Psi^{(A)}_0 \rangle = 
\sum_\beta c_{\alpha\beta} \phi_\beta(\bm{r})
\end{equation}
where $\Psi$ is a many-body state of either the $A$ or $A-1$ nucleus, 
$\alpha$ is a state label with 0 being the target ground state,
$c_{\alpha\beta}$ is a parentage coefficient, and 
$\phi_\beta$ is a local overlap function that is expected to 
resemble an orbital for the mean field.
If the expansion basis is orthnormal, the spectroscopic factors
become
\begin{equation}
S_\beta = \sum_\alpha \left| c_{\alpha\beta} \right|^2
\end{equation}
The parentage expansion depends upon the nuclear structure and
is independent of the probe; hence, the spectroscopic factors
should not depend upon $Q^2$.
However, the assumption that the electromagnetic current
operator is adequately approximated by the free nucleon current
with the energy transfer placed on shell and with free form 
factors is questionable.
Such approximations violate current conservation, Lorentz 
covariance, and unitarity.
Furthermore, the current in a many-body system becomes nonlocal 
when expressed in terms of single-nucleon degrees of freedom;
such nonlocality could contribute to an apparent scale dependence.
Nevertheless, one must resort to simplified models of the
current operator because there is no practical method for
performing rigorous calculations for nucleon knockout from 
nuclei with $A>2$.
Therefore, $Q^2$ dependent modifications of the off-shell 
single-nucleon current operator should not be surprising.

The recent JLab data for $^{12}$C show no evidence for 
systematic variations of the current operator within the
range $0.6 \leq Q^2 \leq 1.8$ (GeV/$c$)$^2$.
Although consistency among previous experiments is not
entirely satisfactory, our RDWIA analysis shows much
less variation with $Q^2$ than suggested by Lapik\'as \etal
Because the continuum contamination for $E_m > 30$ MeV
has not been modeled accurately, we consider $N_{1p}$ for
the $p$-shell bin to be the most reliable gauge of 
possible $Q^2$ dependencies.
Comparing the average of the two Saclay results for
quasiperpendicular kinematics with the average of the
JLab results for NLSH and EDAD1, we see an increase
of $0.87/0.69 = 1.26$ compared with a factor of about
1.6 for $S_{1p}+S_{1s}$ from Ref. \cite{Lapikas00}.
There may still be a small effect, which we would
attribute to the current instead of spectroscopic factors,
but until the failure of RDWIA to reproduce low $Q^2$ 
data for parallel kinemtics is understood we are
reluctant to interpret this as a simple scale factor.

\subsection{Transparency}
\label{sec:transparency}

The experimental definition of nuclear transparency 
\begin{equation}
\label{eq:transp}
{\cal T}_\text{exp} = \frac{ \int_V d^3p_m dE_m d\Omega_e \; 
\sigma_{e,e^\prime p}}
{f \sum_\alpha \int_V d^3p_m dE_m d\Omega_e \; K
\rho_\alpha(p_m) S_\alpha(E_m) \sigma_{ep}}
\end{equation}
compares the measured semi-inclusive differential cross section for proton 
knockout with a PWIA calculation based upon factorization of a model spectral 
function and the free {\it e-p} cross section.
The phase-space volume is indicated by $V$ and we leave possible correction 
factors for nonuniform acceptance implicit, assuming that they are applied in 
both numerator and denominator consistently.
Here $K$ is a kinematic factor, $\rho_\alpha(p_m)$ is the momentum distribution
for orbital $\alpha$, 
$S_\alpha(E_m)$ is the energy distribution for single-nucleon removal from
orbital $\alpha$, and $f$ is a correction factor describing the depletion of
the single-hole spectral function by correlations that shift strength to large 
$E_m$.
Dutta \etal assumed that $f$ is independent of $\alpha$ and obtained an 
estimate of 0.9 for $^{12}$C from \cite{Makins94} that
is consistent with the $1p$ normalization factor we extracted in the 
preceding section.

Similarly, a theoretical definition of nuclear transparency appropriate to
quasiperpendicular kinematics takes the form
\begin{equation}
\label{eq:Tperp}
{\cal T}_\bot =
\frac{ \sum_\alpha \int dp_m p_m \; \sigma_\text{RDWIA}(p_m,E_\alpha)}
{\sum_\alpha \int dp_m p_m \; \sigma_\text{RPWIA}(p_m,E_\alpha)}
\end{equation}
where the numerator integrates the RDWIA and the denominator the RPWIA 
cross section for the same model of the spectral function.
For simplicity we neglect the spread of electron kinematics and assume 
quasifree kinematics for the ejectile, neglecting the small variation of 
transparency with nucleon kinetic energy within the acceptance.
Similarly, we also neglect variations over the spreading widths of 
nuclear orbitals and assume that the bound-nucleon kinematics can be 
approximated by the IPSM.
A more rigorous calculation would use Monte Carlo integration over a realistic
model of the experimental acceptance, but the dependence of nuclear 
transparency upon energy is very mild and the differences between 
transparencies for parallel and quasiperpendicular kinematics are 
small \cite{Kelly96b}.

Note that one should not include the correlation factor $f$ in the theoretical
definition of transparency because any strength that is removed from the 
numerator by correlations is also removed from the denominator by using
the same spectral function for both.
This factor is appropriate for Eq. (\ref{eq:transp}) because the numerator
integrates experimental yield while the denominator integrates a theoretical
calculation based upon a model spectral function.
When the denominator uses the IPSM one requires a theoretical estimate for 
the fraction of the the spectral strength that is shifted out of the 
experimental range of $E_m$. 
Unfortunately, a recent comparison of the data with RDWIA and Glauber 
calculations includes the correlation factor in both theoretical and 
experimental definitions of transparency \cite{Lava04}, 
which we believe is incorrect.

\begin{figure}
\centering
\includegraphics[width=4in]{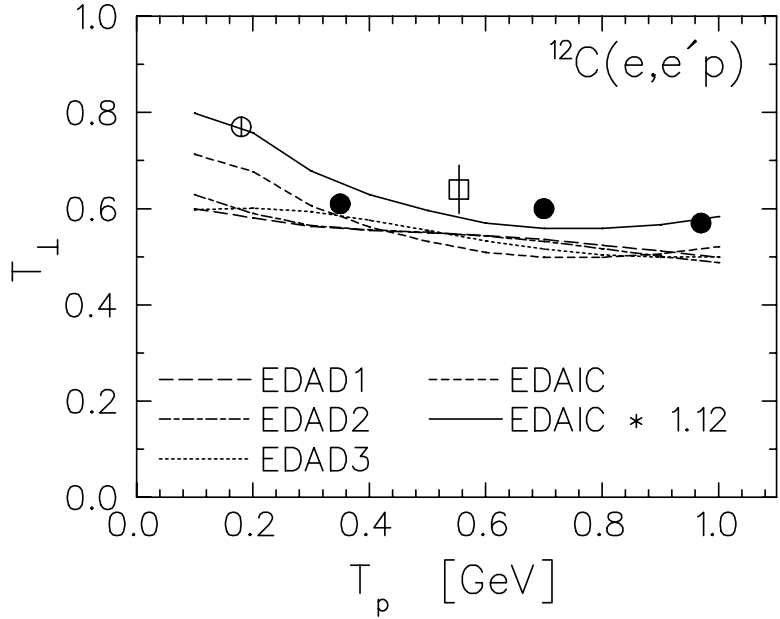}
\caption{
RDWIA calculations using several optical potentials are compared to 
transparency data for $^{12}$C$(e,e^\prime p)$.
The solid curve has been multiplied by 1.12 to provide a better fit to 
the data from MIT (open circle), SLAC (open square) and JLab (filled circles.
The SLAC datum includes systematic uncertainties, but other error bars
are statistical only.}
\label{fig:transparency}
\end{figure}

RDWIA calculations using NLSH wave functions and several optical potentials 
from Dirac phenomenology \cite{Cooper93} are compared in 
Fig. \ref{fig:transparency} with transparency data for 
$^{12}$C$(e,e^\prime p)$ from MIT \cite{Garino92}, 
SLAC \cite{Makins94,O'Neill95}, and JLab \cite{Abbott98,Dutta03}.
The present results are very similar to those we previously obtained 
\cite{Kelly96b} using the EMA-noSV approximation and a definition of 
transparency that replaces the differential cross sections in 
Eqs. (\ref{eq:transp},\ref{eq:Tperp})
with the corresponding reduced cross sections.
Neither change is more than a couple percent.
Our results are also similar to the RDWIA results of Lava \etal \cite{Lava04}
after removal of the correlation factor.
Also note that Lava \etal used a transparency definition based upon
reduced cross section.
However, Meucci \cite{Meucci02a} obtained somewhat larger transparency 
factors using an EMA-SV calculation based upon differential cross section.
The latter are the only RDWIA calculations, of which we are aware, that 
slightly overestimate the transparency for $^{12}$C$(e,e^\prime p)$, 
but the origin of their enhancement is not known.

The variation among $A$-independent global optical potentials is small, but
the $A$-dependent potential fit to data for $^{12}$C$(p,p)$ provides
the best description of the energy dependence for $T_p \lesssim 400$ MeV.
However, all of these calculations systematically underestimate the 
experimental transparency.
Multiplying the EDAIC results by 1.12 provides a reasonable description of 
the data over this range of ejectile energy.
Interestingly, this factor is very close to the inverse of the correlation
correction applied by Abbott \etal \cite{Abbott98}.
If we assume that their correction factor is a reasonable estimate of the
fraction of the single-nucleon yield shifted by correlations to large $E_m$,
this comparison suggests that the experimental yield is approximately $12\%$
larger than single-nucleon knockout.
This is consistent with the fact that the RDWIA normalization factors sum
to the full IPSM strength without depletion.
Therefore, either the RDWIA is approximately $12\%$ too absorptive or there
is a significant continuum arising from multinucleon knockout that contributes
approximately $12\%$ of the strength for $E_m < 80$ MeV.
The rather strong $^{16}$O$(e,e^\prime p)$ continuum seen in Fig. 21 of 
Ref. \cite{Fissum04} suggests that the underlying continuum is probably
responsible for the large normalization factors and transparency values 
for $^{12}$C$(e,e^\prime p)$.

The consistency between the fitted $1p$ normalization factors for
$Q^2 > 0.6$ (GeV/$c$)$^2$ and the predicted IPSM depletion factor supports 
the accuracy of $(e,e^\prime p)$ attenuation factors calculated using RDWIA 
with global optical potentials.
The observation that the yield for $30 \leq E_m \leq 80$ MeV is considerably
larger than can be attributed to $1s$ knockout demonstrates that there
is significant background in the numerator that is not described by
the spectral function used in the denominator of Eq. (\ref{eq:transp}).
The contribution to this background from multinucleon currents artificially
increases the measured transparency insofar as that quantity is intended
to represent the loss of flux from single-nucleon knockout channel.
However, the contribution to this background from rescattering processes
that simply redistribute that flux within the experimental acceptance,
instead of removing it entirely, represents a limitation of the 
distorted wave approximation which does not account for where flux
``absorbed'' by the optical potential finally appears \cite{Kelly96b}.
The background fraction probably increases with $A$, but a more detailed
model of the continuum is needed to distinguish between the processes that
increase the experimental transparency with respect to RDWIA.

\section{Summary and Conclusions}
\label{sec:conclusions}

We have used RDWIA calculations based upon Dirac-Hartree bound-state wave 
functions to analyze data for $^{12}$C$(e,e^\prime p)$ for $E_m < 80$ MeV.
Good fits are obtained for the recent JLab data \cite{Dutta03} for 
$0.6 \leq Q^2 \leq 1.8$ (GeV/$c$)$^2$, with a slight preference for NLSH 
over HS wave functions and practically no variation with any of the
optical potentials provided by Cooper \etal \cite{Cooper93}.
We find that the $p$-shell bin contains approximately 87\% of the IPSM
strength independent of $Q^2$ over that range.
Although the $s$-shell bin appears to carry almost 100\% of IPSM, 
independent of $Q^2$, there is evidence in the momentum distribution for 
a significant continuum that artificially increases the $1s$ normalization 
factor for two-component fits mixing $1p_{3/2}$ and $1s_{1/2}$ contributions.
Therefore, we consider the $1p$ contribution to the $15 \leq E_m \leq 25$ 
MeV bin to be the most reliable estimate of the depletion of IPSM 
orbitals by short-range correlations; the occupancy of 87\% is consistent
with estimates based upon the self-consistent Green's function method
and with recent direct measurements of the correlated continuum
using parallel kinematics \cite{Rohe04,Muther04}.

We have also analyzed the same low $Q^2$ data sets used by 
Lapikas \etal \cite{Lapikas00} to study the $Q^2$ dependence of 
normalization factors for $^{12}$C$(e,e^\prime p)$.
We find that the RDWIA calculations for Dirac-Hartree wave functions
reproduce the low $Q^2$ data for quasiperpendicular kinematics well,
but with somewhat smaller normalization factors.
For example, the $1p$ normalization factors for two Saclay experiments
were 0.63 and 0.75 using NLSH wave functions and EDAD1 optical potentials,
compared with 0.87 for the JLab data.
However, the same model fails to reproduce low $Q^2$ data for parallel
kinematics and remediation of this problem will require more than a 
multiplicative factor.
Similar problems in previous NRDWIA analyses prompted van der Steenhoven
\etal \cite{vdSteenhoven88a} to adjust the Woods-Saxon wave function, 
the optical potential, and an empirical enhancement of the transverse 
components of the current simultaneously.
Although a better fit was achieved, this wave function does not 
reproduce the data for quasiperpendicular kinematics for either 
early low $Q^2$ experiments or the more recent high $Q^2$ experiments.
The momentum distributions for the fitted Woods-Saxon wave functions are 
too narrow, which artificially reduces the normalization factors fit
to data.
Thus, Lapikas \etal \cite{Lapikas00} obtained a $1p$ normalization factor
of only 0.56 for $Q^2 < 0.2$ (GeV/$c$)$^2$ and suggested that spectroscopic
factors might depend strongly upon the resolution of the probe.
We argue that spectroscopic factors are properties of nuclear structure 
that are independent of probe, but that the effective current operator
for single-nucleon knockout may include additional $Q^2$ dependencies
beyond those in familiar off-shell extrapolations of the free nucleon
current operator. 
On the other hand, the additional $Q^2$ dependence is probably smaller 
than their estimate.
Further investigation is needed to explain the discrepancy between 
parallel and quasiperpendicular kinematics at low $Q^2$  --- we believe
that it is unwise to fit the momentum distribution to data with a 
strong variation of $x=Q^2/2m\omega$ in the absence of a more
fundamental theory of the effective current operator.

Finally, we found that RDWIA calculations of nuclear transparency need
to be multiplied by approximately 1.12 to reproduce data for $^{12}$C 
and attribute much of this enhanced transparency to contributions to the
continuum for $30 \leq E_m \leq 80$ MeV that are not directly related
to single-nucleon knockout, such as multinucleon electromagnetic currents.
However, a more detailed model of the continuum is need to distinguish 
between contributions due to multinucleon currents and rescattering
processes.

\begin{acknowledgments}
We thank Dr. Dutta for data tables and Dr. Ud\'ias for tables of the 
NLSH and HS wave functions.  
We especially thank Dr. Lapik\'as for the parameters of the NIKHEF wave 
functions and data tables for the Tokyo, Saclay, and SLAC experiments.
The support of the U.S. National Science Foundation under grant PHY-0140010
is gratefully acknowledged.
\end{acknowledgments}


\end{document}